\documentclass[aps,prc,twocolumn,groupedaddress,showpacs]{revtex4}
\usepackage{graphicx}

\begin{document}


\title{Electromagnetic dissociation of $^{8}$B and the astrophysical S factor for
$^{7}$Be($p,\gamma)^{8}$B}

\author{B. Davids}
\email[]{davids@triumf.ca}
\affiliation{Kernfysisch Versneller Instituut, Zernikelaan 25, 9747 AA Groningen, The Netherlands}
\affiliation{TRIUMF, 4004 Wesbrook Mall, Vancouver BC V6T 2A3, Canada}
\author{S. Typel}
\email[]{s.typel@gsi.de}
\affiliation{Gesellschaft f\"ur Schwerionenforschung mbH, Planckstr. 1, 64291 Darmstadt, Germany}

\date{\today}

\begin{abstract}
We present S factor data obtained from the Coulomb dissociation of 83 MeV/nucleon $^8$B, and analyze $^7$Be longitudinal momentum distributions measured at 44 and 81 MeV/nucleon using a potential model, first-order perturbation theory, and dynamical solution of the time-dependent Schr\"odinger equation. Comparing our results with independent continuum-discretized coupled channels calculations, we study the reaction model and beam energy dependence of the $E2$ contribution to the dissociation cross section. By fitting radiative capture and Coulomb breakup data taken below relative energies of 400 keV with potential models constrained by $^7$Li~+~$n$ and $^7$Be~+~$p$ elastic scattering data, we examine the mutual consistency of recent S$_{17}$ measurements and obtain a recommended value for S$_{17}$(0) of $18.6 \pm 0.4$~(experimental)~$\pm$~1.1~(extrapolation)~eV~b ($1\sigma$). This result is in good agreement with recent experimental determinations of the asymptotic normalization coefficient of the valence proton wave function in $^8$B.
\end{abstract}


\maketitle

\section{Introduction}

With the discovery that a large fraction of the high-energy electron neutrinos emitted in the $\beta^+$ decay of  $^8$B in the sun are transformed into other active neutrino flavors on their way to terrestrial detectors \cite{ahmad02}, and, consistent with CPT symmetry, that reactor-produced antineutrinos also oscillate \cite{eguchi02}, a robust solution of solar neutrino problem seems to be in hand. Although the SNO neutral current measurement of the $^8$B solar neutrino flux is consistent with the predictions of the standard solar model \cite{bahcall01}, its precision is still far from that required to ascertain if there is transformation of solar electron neutrinos into sterile neutrinos \cite{barger02, bahcall02}. In order to reach this goal, the precision of both the experimental measurement and the theoretical prediction must be improved. Through the addition of NaCl and $^3$He proportional counters to the SNO detector, its neutral current measurement will be made much more precise. In order to improve the precision of the theoretical prediction of the $^8$B neutrino flux, the rate of the reaction that produces $^8$B in the sun, $^{7}$Be($p,\gamma)^{8}$B, must be better determined. Currently, the uncertainty on the zero-energy astrophysical S factor for this reaction, S$_{17}$(0) makes the largest contribution to the theoretical error budget.

Despite the profusion of recent determinations of S$_{17}$(0) \cite{kikuchi98,hass99,iwasa99,strieder01,terrasi01,azhari01,davids01,hammache01,trache01,junghans02,baby03,baby03a,schuemann03}, the current recommendation \cite{adelberger98} is based on the results of a single experimental measurement \cite{filippone83}. Recently, a weighted mean of such determinations based on both direct and indirect methods was given \cite{davids01a}. Here, we present data obtained via the indirect method of Coulomb dissociation \cite{baur86}, offer theoretical analyses of these and other data, and assess the current state of knowledge of S$_{17}$(0).

At solar energies, the radiative capture $^{7}$Be($p,\gamma)^{8}$B is entirely $E1$ dominated. The Coulomb dissociation of $^8$B at intermediate beam energies also proceeds predominantly via $E1$ transitions, but the much larger flux of $E2$ virtual photons leads to a finite $E2$ contribution to the dissociation cross section. This complicates efforts to extract the $E1$ strength that determines the radiative capture rate. Since the first measurement of the Coulomb dissociation of $^8$B \cite{motobayashi94}, the role of $E2$ transitions in this process has been rather controversial. Experimental efforts to determine the $E2$ contribution to the breakup cross section have focused on the relative energy and $^8$B scattering angle distributions \cite{kikuchi97,iwasa99}, the angular distributions \cite{schuemann03} and longitudinal momentum distributions of the emitted protons \cite{davids01}, and the longitudinal momentum distributions of the $^7$Be fragments \cite{kelley96,davids98}. The relative energy and $^8$B scattering angle distributions are relatively insensitive to $E2$ strength, since the $E1$ and $E2$ contributions sum incoherently \cite{bertulani03}. In contrast, the fragment longitudinal momentum and angular distributions are better suited to this task because interference between $E1$ and $E2$ transition amplitudes produces striking asymmetries in these distributions \cite{esbensen96}.

\section{Longitudinal Momentum Distributions and $E2$ Strength}

A first-order perturbation theory analysis of $^7$Be longitudinal momentum distributions found that the $E2$ matrix elements implicit in the potential model used had to be scaled by 0.7 in order to best fit the measured distributions \cite{davids01a}. In that work, the notion that higher-order dynamical effects quench the $E2$ strength predicted by first-order theories \cite{esbensen96} was advocated. Support for this idea was provided by concurrent \cite{davids01a} and subsequent \cite{mortimer02} continuum-discretized coupled channels (CDCC) calculations. The perturbative calculations \cite{esbensen96} employ a simple, single-particle model for the structure of $^8$B, and calculate the electromagnetic dissociation cross section assuming single photon exchange, neglecting nuclear-induced breakup. The CDCC calculations use a similar structure model but include electromagnetic and nuclear matrix elements to all orders. In Ref.\ \cite{mortimer02}, the authors found that a scaling of the $E2$ matrix elements by a factor of 1.6 was required for the best description of the data using CDCC. Hence the $E2$ strength had to be reduced in first-order calculations, and enhanced in all-order calculations, relative to the original structure model predictions.

Here we study the same longitudinal momentum distributions \cite{davids01a} using two approaches, first-order perturbation theory and dynamical solution of the time-dependent Schr\"odinger equation \cite{typel99}. Both methods employ only Coulomb matrix elements, and assume the same single-particle potential model. In this model, the $p$-wave potential depth has been fixed by the $^{8}$B binding energy, and the depths of the potential for all other partial waves have been fixed by the $s$-wave scattering lengths for channel spin 1 and 2 in the isospin mirror $^7$Li$~+~n$ system \cite{koester83}. The potential radius is 2.5 fm, the diffuseness 0.65 fm, and the $p$-wave potential depth is 43.183 MeV. For $s$, $d$, and $f$-waves, the potential depths are 43.857 and 52.597 MeV for $S$~=~1 and $S$~=~2 respectively. Spectroscopic factors for the two spin configurations were taken from Ref.\ \cite{cohen67} after a center-of-mass correction, and are 0.3231 for $S$~=~1 and 0.8572 for $S$~=~2. These theoretical spectroscopic factors agree very well with other shell model calculations \cite{brown96} and cluster model calculations \cite{csoto00}. Our potential model and that used in Ref.\ \cite{mortimer02} differ from that of Ref.\ \cite{esbensen96} in that they lack a spin-orbit interaction. Accordingly, our potential model includes no $1^+$ resonance at 0.6 MeV and lacks resonant $E2$ strength in this region. The present model gives $S_{E2}/S_{E1}$(0.6 MeV) = $5.9\times10^{-4}$, compared with the $9.5\times10^{-4}$ predicted by the model of Ref.\ \cite{esbensen96}. These two values span the range of theoretical predictions made using disparate methods cataloged in Ref.\ \cite{davids01a}.

We have scaled the $E2$ matrix elements calculated with our potential model, and then performed perturbative and dynamical calculations with the different $E2$ scaling factors. In the dynamical calculations the wave function of the $^7$Be~+~$p$ system was expanded into partial waves, taking relative orbital angular momenta $\ell\leq3$ into account. The radial wave function for each partial wave, specified by $\ell$ and its projection $m$, was discretized on a mesh with exponentially increasing step size covering radii between 0 and 900 fm. Starting with the ground state of $^8$B, the evolution of the coupled radial wave functions was followed by applying a unitary approximation of the time-evolution operator in steps of 5 fm/c. A smooth switch-on and switch-off of the Coulomb potential for large distances was introduced to avoid spurious excitations. Only multipoles $\lambda$ = 1 and 2 were considered in the expansion of the time-dependent Coulomb interaction while the $^8$B center of mass followed a hyperbolic Coulomb trajectory with respect to the target. The final time-evolved wave function was projected onto the  relevant continuum wave functions to obtain the excitation amplitudes and the cross sections. Allowing the overall normalization to vary freely and convoluting the perturbative and dynamical calculations with the experimental momentum resolution of 5 MeV/c, we have determined the $E2$ scaling factor that minimizes the $\chi^2$ value for the central six 44 MeV/nucleon data points and the central five 81 MeV/nucleon data points of the measured longitudinal momentum distributions. 

\begin{figure}\includegraphics[width=\linewidth]{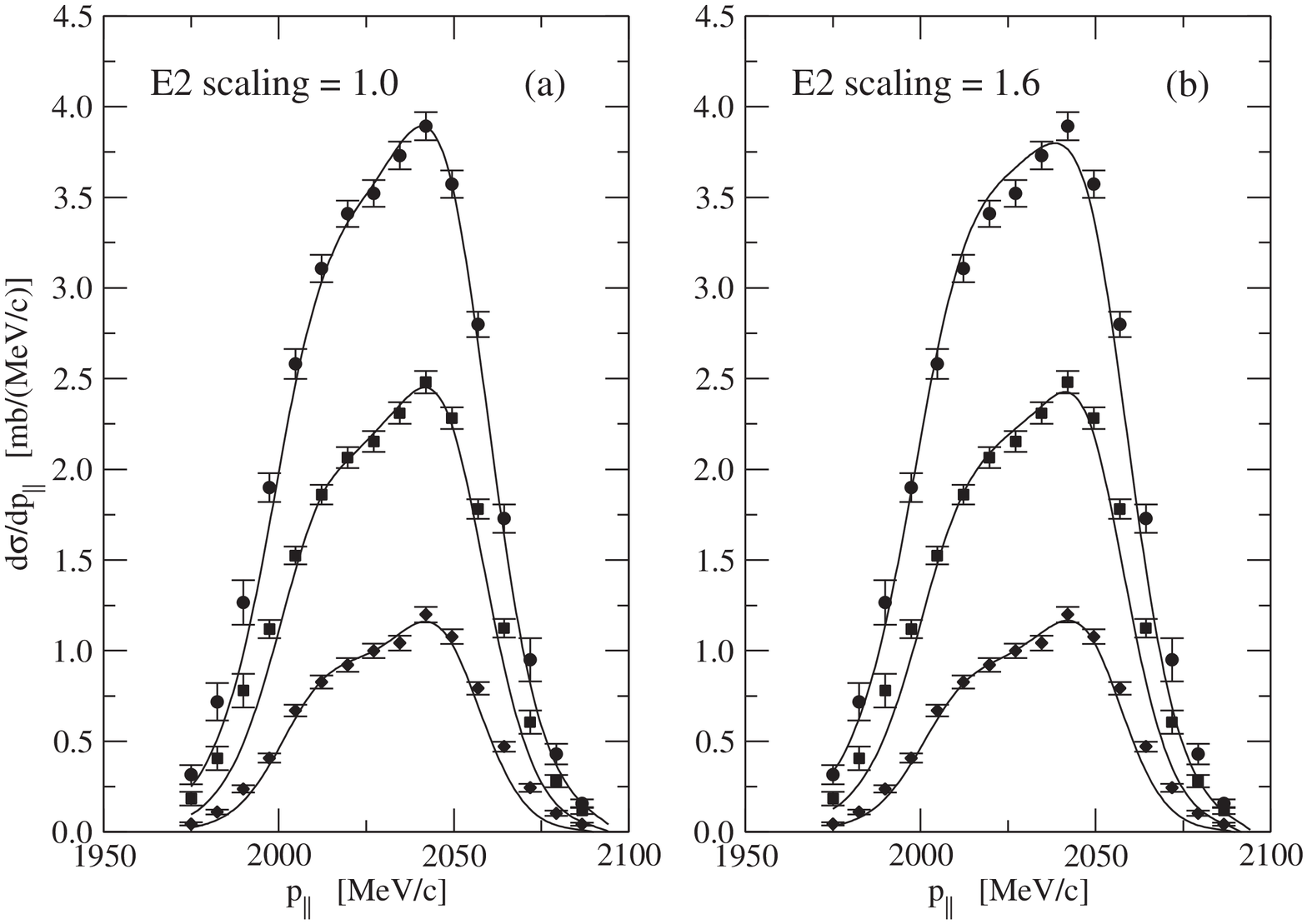} \caption{Measured $^7$Be longitudinal momentum distributions from the Coulomb dissociation of 44 MeV/nucleon $^{8}$B on Pb with
$^7$Be scattering angle cuts of 1.5$^\circ$, 2.4$^\circ$, and 3.5$^\circ$. Only relative errors are shown. The solid curves in panel (a) are the results of first-order perturbation theory calculations, while panel (b) shows dynamical calculations. In both cases the $E2$ matrix elements have been scaled by the indicated factor.} \label{fig1} \end{figure}

\begin{figure}\includegraphics[width=\linewidth]{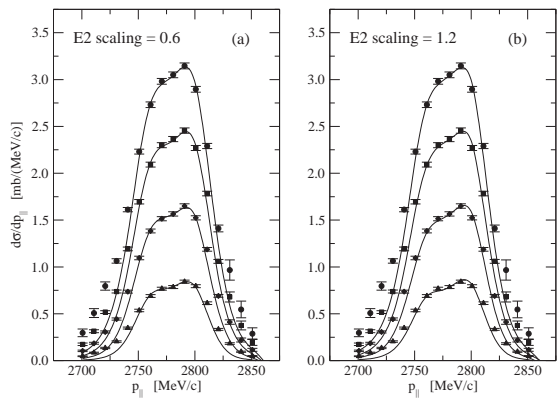} \caption{Same as Fig.\ \ref{fig1}, but for 81 MeV/nucleon $^{8}$B on Pb with $^7$Be scattering angle cuts of 1.0$^\circ$, 1.5$^\circ$, 2.0$^\circ$, and 2.5$^\circ$.} \label{fig2} \end{figure}

Fig.\ \ref{fig1} shows the results of the best-fit calculations for the breakup of $^8$B on Pb at 44 MeV/nucleon, while Fig.\ \ref{fig2} shows the 81 MeV/nucleon results. Both the perturbative and the dynamical calculations satisfactorily reproduce the measured asymmetries in the central regions of the longitudinal momentum distributions, though neither calculation describes the widths of the 81 MeV/nucleon distributions very well; the measured distributions are wider. As the tails of the momentum distributions correspond to high $^8$B excitation energies, it is possible that our neglect of the nuclear interaction, which is relatively more important at high excitation energies than low, is responsible for this discrepancy. The required overall normalization factors range from 0.7 to 0.95 for the various angle cuts, beam energies, and reaction models. Consistent with earlier findings \cite{esbensen96,davids01a,mortimer02,esbensen02}, higher-order effects present in the dynamical calculations but neglected in the first-order perturbative calculations tend to reduce the asymmetry predicted for a given $E2$ strength. Hence smaller $E2$ scaling factors are required for the perturbative calculations than for the dynamical calculations. In addition, there is a hint of a beam energy dependence apart from that inherent in standard Coulomb excitation theory, with less $E2$ strength required to explain the asymmetries observed at the higher beam energy than at the lower. Such an effect, which has thus far eluded theoretical explanation, may be responsible for the reported lack of evidence for $E2$ contributions to Coulomb breakup at 254 MeV/nucleon \cite{iwasa99,schuemann03}.

The best-fit $E2$ matrix element scaling factors and their 1$\sigma$ uncertainties obtained using the two reaction models are shown in Table \ref{table1}. It is of interest to compare the results found here with those reported in Ref.\ \cite{davids01a}, which were deduced from a perturbative analysis of the 3.5$^\circ$ 44 MeV/nucleon longitudinal momentum distribution. The best-fit scaling factor obtained from comparison of our perturbative calculation with the same data here implies $S_{E2}/S_{E1}$(0.6 MeV) = $5.9^{+2.6}_{-2.1}\times10^{-4}$, which is consistent with the value of $4.7^{+2.0}_{-1.3}\times10^{-4}$ given in Ref.\ \cite{davids01a}. From this we conclude that the results of perturbative analyses favored thus far by workers in the field are relatively insensitive to the details of the underlying structure model. This analysis corroborates the case for including $E2$ contributions in theoretical descriptions of the Coulomb breakup of $^8$B at beam energies below 100 MeV/nucleon.

\begin{table}
 \caption{Best-fit scaling factors of $E2$ matrix elements predicted by our potential model obtained from fitting the longitudinal momentum distributions of $^7$Be fragments from the Coulomb dissociation of $^8$B on Pb. The uncertainties are 1$\sigma$ values.\label{table1}}
 \begin{ruledtabular}
 \begin{tabular}{ccc}
Beam Energy (MeV/nucleon)&Calculation&Scaling Factor\\ \hline
44&perturbative&$1.01\pm0.21$\\
44&dynamical&$1.6\pm0.2$\\
81&perturbative&$0.63\pm0.12$\\
81&dynamical&$1.2\pm0.2$\\
 \end{tabular}
\end{ruledtabular}
 \end{table}
 
One can also compare the results obtained here with dynamical calculations to those found using CDCC calculations in Ref.\ \cite{mortimer02}. Those authors studied both the 44 MeV/nucleon and 81 MeV/nucleon longitudinal momentum distributions. Although no uncertainty was specified, the best-fit scaling value reported in Ref.\ \cite{mortimer02} is 1.6, in perfect agreement with the value of 1.6 $\pm$ 0.2 found from the dynamical analysis of 44 MeV/nucleon data here. Our result for the 81 MeV/nucleon data is somewhat smaller, but if the CDCC scaling factor uncertainty is comparable to that found here, the two results agree reasonably well at this beam energy also. Moreover, the fact that nuclear matrix elements are included in the CDCC calculations but not in our dynamical calculations suggests that nuclear-induced breakup neither significantly enhances nor reduces the asymmetries of the Coulomb dissociation calculations for the large impact parameters considered here. Since the potential model assumed in the CDCC calculations is very similar to that employed here, we conclude that these two very different methods predict the same influence of higher-order effects. We therefore consider these predictions to be robust, and conclude that the practice of scaling the $E2$ matrix elements in first-order perturbative calculations is a reasonable and practical way to proceed with the analysis of $^8$B Coulomb dissociation experiments.

\section{Dissociation Cross Section and S$_{17}$}
 
In Ref.\ \cite{davids01}, an exclusive measurement of the Coulomb dissociation of 83 MeV/u $^{8}$B on a Pb target was described and experimental data were presented in graphical form, including the Coulomb dissociation cross section for $^{8}$B scattering angles $\leq 1.77^{\circ}$. This experimentally measured cross section was interpreted in the context of first-order perturbation theory in order to obtain S$_{17}$(0). Here we give the energy-dependent S factor derived from these data in both graphical and tabular form. Table\ \ref{table2} contains the measured cross sections and their total 1$\sigma$ uncertainties. The systematic uncertainty common to each point amounts to 7.1\%, and includes contributions from the target thickness (1\%), beam intensity (2.6\%), momentum calibration (4.2\%), and the size of the $E2$ component (5\%).

\begin{table}
 \caption{Experimentally measured cross section for the Coulomb dissociation of 83 MeV/u $^8$B on Pb for $^8$B laboratory scattering angles of 1.77$^{\circ}$ and less (impact parameters $\geq$ 30 fm), and astrophysical S factor for the $^{7}$Be($p,\gamma)^{8}$B reaction. Uncertainties are 1$\sigma$ values, and the systematic uncertainty common to each point is 7.1\%. The relative energy bins over which the cross sections and S factors have been averaged extend to the midpoints of the intervals separating adjacent data points.\label{table2}}
 \begin{ruledtabular}
 \begin{tabular}{ccc}
E$_{rel}$ (MeV)& d$\sigma$/dE$_{rel}$ (mb/MeV) & S$_{17}$(E$_{rel}$) (eV b)\\ \hline
0.064 & 9.0 $\pm$ 2.8 &  \\
0.192 & 50.4 $\pm$ 4.9 & 15.2 $\pm$ 1.5\\
0.384 & 113.2 $\pm$ 10.1 & 16.1 $\pm$ 1.4\\
0.768 & 116.1 $\pm$ 10.9 & 19.3 $\pm$ 1.8\\
1.280 & 61.2 $\pm$ 6.8 & 23.3 $\pm$ 2.6\\
1.792 & 41.9 $\pm$ 5.9 & 35.2 $\pm$ 5.0\\
 \end{tabular}
\end{ruledtabular}
 \end{table}

Astrophysical S factors derived from the experimental cross section data are shown in Fig.\ \ref{fig4} and Table\ \ref{table2}. These S factors have been corrected for the $E2$ contribution to the breakup cross section in the following way. The measured cross section was multiplied by the fraction of the total electromagnetic dissociation cross section attributable to $E1$ transitions to obtain the $E1$ cross section. This $E1$ fraction was calculated using the Esbensen and Bertsch structure model \cite{esbensen96} and first-order perturbation theory, calibrated by comparison with the measured longitudinal momentum distributions as described in Ref.\ \cite{davids01a}. $M1$ transitions were included by folding the S factor measured at the $1^+$ resonance by Filippone {\it et al.} \cite{filippone83} with the calculated virtual photon spectrum. The energy-dependent S factor was deduced from the $E1$ cross section using semiclassical first-order Coulomb excitation theory \cite{alder56}, taking into account the experimental relative energy resolution. A comparison of theoretical predictions of $E2$ strength \cite{davids01a} found that the smallest prediction differed from the largest by less than 40\%. Recent work that examined the influence of structure model assumptions on the calculated $E1$ fraction \cite{forssen03} is consistent with this finding. Since $E2$ transitions account for only 5\% of the measured cross section in the perturbative analysis applied to the data of Ref.\ \cite{davids01}, the 5\% uncertainty attributed here to the $E2$ component is quite conservative. Owing to our own findings regarding the size of higher-order dynamical effects at very low relative energies as well as those of others \cite{alt03}, we have not extracted an S factor from the lowest energy data point.

\begin{figure}\includegraphics[width=\linewidth]{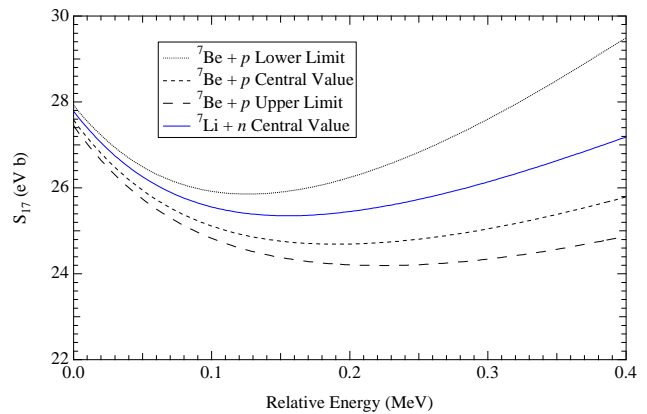} \caption{(Color online) Calculated astrophysical S factors for the $^{7}$Be($p,\gamma)^{8}$B reaction based on potential models constrained by $^7$Be~+~$p$ and $^7$Li~+~$n$ elastic scattering data. The 1$\sigma$ upper and lower limits on the measured scattering lengths in the $^7$Be~+~$p$ system were used to construct parameter sets that yield the curves labeled as upper and lower limits. The corresponding curves for the $^7$Li~+~$n$ case are indistinguishable from that based on the central values of these more precisely measured scattering lengths.} \label{fig3} \end{figure}

\begin{figure}\includegraphics[width=\linewidth]{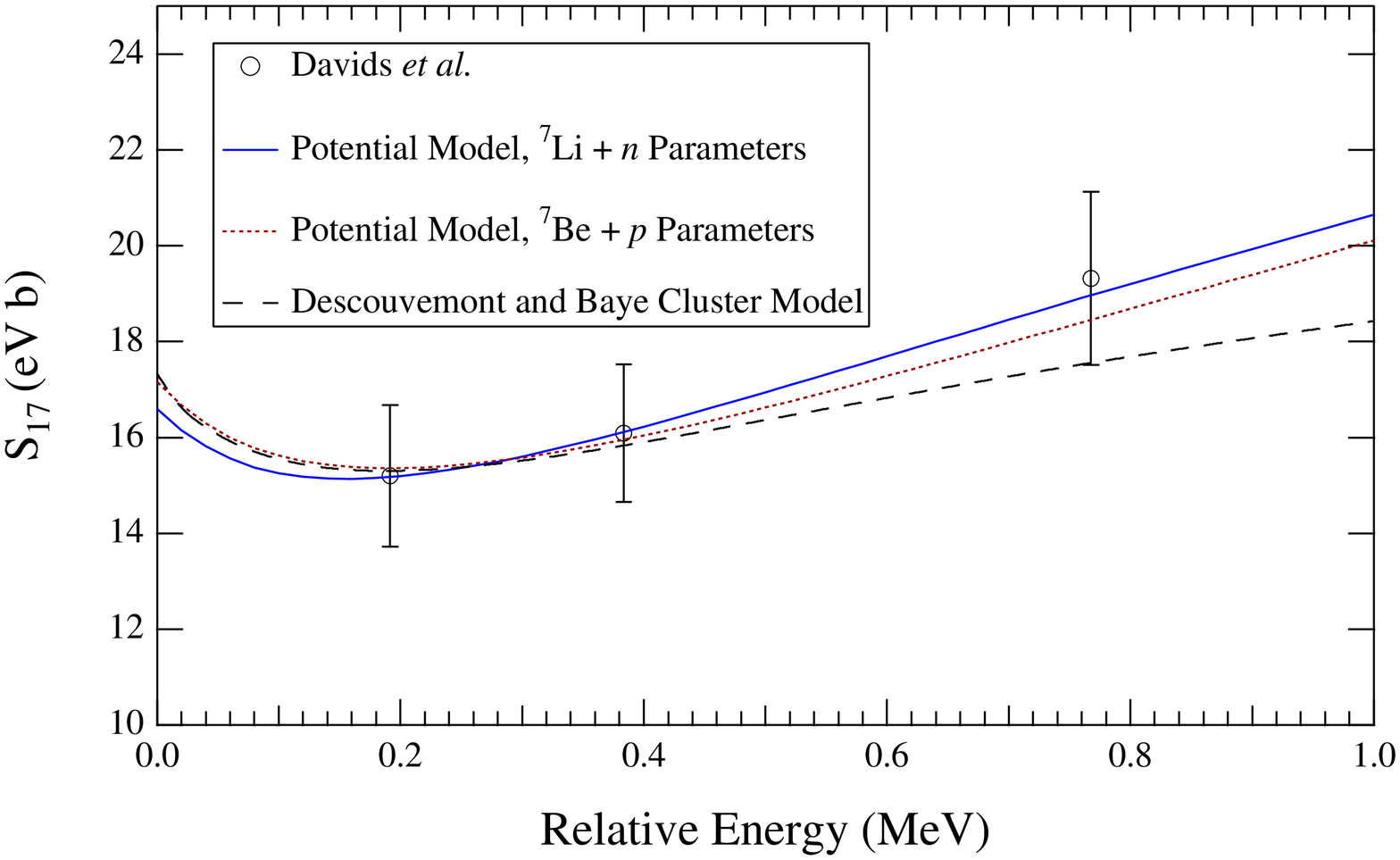} \caption{(Color online) Astrophysical S factor for the $^{7}$Be($p,\gamma)^{8}$B reaction obtained from the Coulomb dissociation of 83 MeV/u $^8$B on Pb. Potential model calculations using parameters fixed by $^7$Be~+~$p$ and $^7$Li~+~$n$ elastic scattering data and a cluster model calculation of Descouvemont and Baye using the Volkov II interaction are also shown. The calculations have been scaled to best fit the data below 400 keV.} \label{fig4} \end{figure}

In order to minimize the uncertainties due to nuclear structure in extrapolating the astrophysical S factor to zero energy, it has been recommended that only data obtained below 400 keV be considered \cite{jennings98}. This conclusion is supported by our own potential model calculations, which are shown in Fig.\ \ref{fig3}. In addition to the potential model calculation described above, we have performed calculations with the same spectroscopic factors but with the $s$, $d$, and $f$-wave potential depths adjusted to reproduce the scattering lengths measured in the $^7$Be~+~$p$ system \cite{angulo03}. The potential depths for the central values of the measured scattering lengths in this case are 24.283 MeV and 51.898 MeV for $S$~=~1 and $S$~=~2 respectively. The scattering lengths  for the $^7$Be~+~$p$ system are not as precisely known as those for the mirror $^7$Li~+~$n$ system; the measured $^7$Be~+~$p$ and $^7$Li~+~$n$ scattering lengths for the dominant $S$~=~2 channel are consistent with isospin symmetry, but there is an inconsistency in the $S$~=~1 channel that requires further study. Since the $^7$Li~+~$n$ scattering lengths have been measured quite precisely, their uncertainties do not significantly affect the shape of the S factor calculated with these potential model parameters. However, the $^7$Be~+~$p$ scattering lengths have large experimental uncertainties which imply a correspondingly large uncertainty in the predicted shape of the S factor when using a potential model tuned to reproduce these elastic scattering data. In Fig.\ \ref{fig3}, we show the results of the potential model calculations with parameter choices that match the measured scattering lengths in the two systems. For the $^7$Be~+~$p$ system, we show the potential calculations corresponding to the central value and 1$\sigma$ upper and lower limits on the measured scattering lengths, while for the $^7$Li~+~$n$ system we only show the central value calculation because the upper and lower limit curves are indistinguishable from it. The extrapolation uncertainty due to the calculated spectroscopic factors is insignificant compared with that due to the uncertainty in potential model parameter choices corresponding to the different scattering lengths.

In Fig.\ \ref{fig4}, the two potential model calculations are shown along with the 83 MeV/nucleon Coulomb breakup S factor data. The solid line is the potential model with $^7$Li~+~$n$ parameters, while the dotted curve is the (central) $^7$Be~+~$p$ parameter set potential model calculation. Both calculations have been scaled to optimally fit the data points below 400 keV, and both reproduce the data well. However, the extrapolated values differ by more than 3\%. This fact exposes the danger of using high relative energy data (above the $M1$ resonance) to extrapolate to zero energy. Here, even limiting the relative energy range to $< 400$ keV and considering only potential models that reproduce the available elastic scattering data, the extrapolation uncertainty is appreciable. The description of these data and those considered in the following section is significantly better with the potential model parameters taken from $^7$Li~+~$n$ rather than $^7$Be~+~$p$ elastic scattering data. Since the $^7$Li~+~$n$ scattering lengths are very precisely measured and isospin is expected to be a good symmetry in this system, we use the potential-model parameters fixed by $^7$Li~+~$n$ elastic scattering data \cite{koester83}, and regard the difference between the extrapolated values using the different potential model parameter sets as the extrapolation uncertainty. It is worth noting that the $^7$Be~+~$p$ scattering lengths are so poorly known that the $^7$Li~+~$n$ potential model curve falls well within the 1$\sigma$ upper and lower limit $^7$Be~+~$p$ scattering length potential model curves shown in Fig.\ \ref{fig3}. Proceeding in this way, we obtain S$_{17}(0) = 16.6 \pm 1.1$ (experimental) $\pm$ 0.9 (extrapolation) eV b. Several recent measurements have been extrapolated to zero energy using a generator coordinate method calculation of Descouvemont and Baye that employed the Volkov II nucleon-nucleon interaction \cite{descouvemont94}. As Fig.\ \ref{fig4} shows, below 400 keV the shapes of this cluster model and the $^7$Be~+~$p$ parameter set potential model are extremely similar; the extrapolated zero-energy S factors for the 83 MeV/nucleon Coulomb breakup data using these two models differ by less than 0.6\%.

\section{Consistency of Experimental S$_{17}$ Determinations}

\begin{figure}\includegraphics[width=\linewidth]{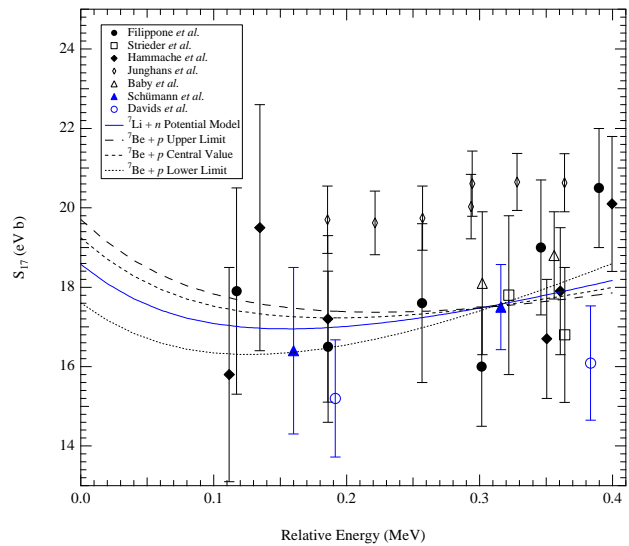} \caption{(Color online) Astrophysical S factors for the $^{7}$Be($p,\gamma)^{8}$B reaction from experimental measurements at relative energies $ < 400$ keV. The potential models constrained by $^7$Be~+~$p$ and $^7$Li~+~$n$ elastic scattering data that best fit all these data except those of Junghans {\it et al.} are also shown. See the text for details.} \label{fig5} \end{figure}

We fit the data of Filippone {\it et al.} \cite{filippone83}, Hammache {\it et al.} \cite{hammache01}, Strieder {\it et al.} \cite{strieder01}, Junghans {\it et al.} \cite{junghans02}, Baby {\it et al.} \cite{baby03}, and Sch\"umann {\it et al.} \cite{schuemann03} using the same procedure. The results of these extrapolations of data below relative energies of 400 keV using our potential models are shown in Table \ref{table3}, and the data themselves appear in Fig.\ \ref{fig5}. The Filippone {\it et al.} data have been renormalized using the weighted average of their two target thickness measurements after adjusting that based on the $^7$Li$(d,p)^8$Li reaction using the cross section recommended in \cite{adelberger98}. They have not been corrected for $^8$B backscattering losses. In order to test the mutual consistency of the seven radiative capture and Coulomb breakup measurements considered here, we have treated them all as a single data set which we fit with the potential models described above. Fitting the 27 data points below 400 keV in this way, we find $\chi^2$ values of 46.5 and 46.6 for 26 degrees of freedom using the $^7$Li~+~$n$ and central $^7$Be~+~$p$ parameter sets respectively. These $\chi^2$ values correspond to a $p$-value of 0.008, indicating inconsistency within the reported uncertainties. Excluding the measurement of Junghans {\it et al.}, which is both very precise and systematically higher than the others, we find $\chi^2$ values of 11.7 and 12.6 for 19 degrees of freedom, corresponding to $p$-values of 0.90 and 0.86. This indicates that the other six measurements are mutually compatible. The best fit potential models for this reduced data set are shown in Fig.\ \ref{fig5}, and the zero-energy S factors for the $^7$Li~+~$n$ and $^7$Be~+~$p$ parameter sets are 18.6(4) and 19.2(4) eV b. Hence apart from a single discrepant measurement, the recent radiative capture and Coulomb breakup measurements below 400 keV can be described consistently by either of our potential model parameter sets.

\begin{table*}
 \caption{Zero energy astrophysical S factors for the $^{7}$Be($p,\gamma)^{8}$B reaction obtained by extrapolating radiative capture and Coulomb dissociation data at E$_{rel}<400$ keV with potential models constrained by $^7$Li~+~$n$ and $^7$Be~+~$p$ elastic scattering data. The experimental uncertainties are $1\sigma$ values.\label{table3}}
 \begin{ruledtabular}
 \begin{tabular}{cccc}
First Author&Reference&S$_{17}$(0) (eV b), $^7$Li~+~$n$&S$_{17}$(0) (eV b), $^7$Be~+~$p$\\ \hline
Filippone&\cite{filippone83}&19.1(8)&19.8(8)\\
Strieder&\cite{strieder01}&18.0(14)&18.7(14)\\
Hammache&\cite{hammache01}&18.8(8)&19.5(8)\\
Junghans&\cite{junghans02}&21.5(3)&22.2(3)\\
Baby&\cite{baby03}&19.5(10)&20.3(10)\\
Sch\"umann&\cite{schuemann03}&18.4(10)&19.0(10)\\
Davids&this work&16.6(11)&17.1(11)\\
All except Junghans&&18.6(4)&19.2(4)\\
 \end{tabular}
\end{ruledtabular}
 \end{table*}

Using the potential model parameters taken from the 1$\sigma$ lower and upper limits on the $^7$Be~+~$p$ scattering lengths, we obtain 17.6(4) and 19.7(4) eV b for S$_{17}$(0) from this reduced data set. Regarding the difference between the zero energy S factors extrapolated with the different potential model parameter sets as the $1\sigma$ extrapolation uncertainty and taking the $^7$Li~+~$n$ parameter set for the central value, we find S$_{17}$(0)~=~18.6~$\pm$~0.4~(experimental)~$\pm$~1.1~(extrapolation) from the six mutually consistent radiative capture and Coulomb breakup measurements.

Recommended values of S$_{17}$(0) based on some of the same experimental data have been offered in previous work \cite{strieder01,junghans02,baby03a}. Our approach differs from others in three principal respects. First, our extrapolation procedure is based on simple potential models constrained by elastic scattering data rather than on a complex cluster model unconstrained by such data. Second, we limit the relative energy range of the experimental data used in the extrapolation to $<$ 400 keV. This is important both because the shapes of the models differ increasingly as relative energy increases, and because the data themselves are less consistent at high relative energies. Extrapolations that include higher relative energy data are dominated by the statistically precise data obtained above the $M1$ resonance, where the uncertainty in the shape of the S factor is problematic. Finally, we evaluate the extrapolation uncertainties by taking the difference between the results of potential models with parameters as extreme as the precision of the elastic scattering data allow, rather than taking the root mean square deviation of a larger set of unconstrained models. Although other groups have limited their extrapolations to the low relative energy regime, we believe that our method has the most solid foundation in experimental data, and will yield both the most reliable central value and a realistic estimate of the extrapolation uncertainty. It is our contention that this extrapolation uncertainty has been universally underestimated, and that a more precise measurement of the $^7$Be~+~$p$ scattering lengths will be required in order to make progress in this area.

Several values of S$_{17}$(0) have been deduced from experimental determinations of the asymptotic normalization coefficient (ANC) of the valence proton wave function in $^8$B, 17.3(18) eV b \cite{azhari01}, 17.4(15) eV b \cite{trache01}, and 17.6(17) eV b \cite{trache03}. A weighted average of these three ANC results gives 17.4(10) eV b. This is somewhat smaller than but perfectly consistent with our central value from the radiative capture and Coulomb breakup data. It is in particularly excellent agreement with the value deduced from the lower limit $^7$Be~+~$p$ parameter set potential model.

\section{Summary}

In summary, we have studied the reaction model and beam energy dependence of the $E2$ contribution to the Coulomb dissociation of $^8$B through the comparison of $^{7}$Be longitudinal momentum distribution data with first-order perturbation theory, continuum-discretized coupled channels, and dynamical calculations. A coherent picture emerges when one considers nine precise direct and indirect measurements of the astrophysical S factor for the $^{7}$Be($p,\gamma)^{8}$B reaction, though one recent measurement is discrepant with the others. By fitting radiative capture and Coulomb breakup data taken below relative energies of 400 keV with potential models constrained by $^7$Li~+~$n$ and $^7$Be~+~$p$ elastic scattering data, we examine the mutual consistency of recent S$_{17}$ measurements and obtain a recommended value for S$_{17}$(0) of $18.6 \pm 0.4$~(experimental)~$\pm$~1.1~(extrapolation)~eV~b ($1\sigma$). This result is in good agreement with recent experimental determinations of the asymptotic normalization coefficient of the valence proton wave function in $^8$B.

\begin{acknowledgments}
BD gratefully acknowledges fruitful discussions with Sam Austin, Carlos Bertulani, Henning Esbensen, Christian Forss\'{e}n, Byron Jennings, Klaus S\"ummerer, and Livius Trache.
\end{acknowledgments}

\bibliography{prc}

\end{document}